\documentclass[pre,aps,twocolumn,superscriptaddress,showpacs]{revtex4-1}

\usepackage{dcolumn}
\usepackage{amsfonts, amssymb, amsmath}
\usepackage{bm}
\usepackage{graphics}
\usepackage{graphicx}
\usepackage{wasysym}
\usepackage{natbib}
\usepackage{epstopdf} 

\usepackage[T1]{fontenc}

\usepackage[misc]{ifsym}

\usepackage[usenames,dvipsnames]{xcolor}

\usepackage{color}
\definecolor{darkblue}{rgb}{0,0,0.6}
\definecolor{darkred}{rgb}{0.6,0,0}

\begin{document}

\title{Jamming and percolation in random sequential adsorption of straight rigid rods on a two-dimensional triangular lattice}

\author{E. J. Perino}
\affiliation{Departamento de F\'{\i}sica, Instituto de F\'{\i}sica Aplicada (INFAP), Universidad Nacional de San Luis - CONICET, Ej\'ercito de Los Andes 950, D5700HHW, San Luis, Argentina}
\author{D. A. Matoz-Fernandez}
\affiliation{Universit\'e Grenoble Alpes, LIPHY, F-38000 Grenoble, France}
\affiliation{CNRS, LIPHY, F-38000 Grenoble, France}
\author{P. M. Pasinetti}
\email[]{mpasi@unsl.edu.ar}
\affiliation{Departamento de F\'{\i}sica, Instituto de F\'{\i}sica Aplicada (INFAP), Universidad Nacional de San Luis - CONICET, Ej\'ercito de Los Andes 950, D5700HHW, San Luis, Argentina}
\author{A.J. Ramirez-Pastor}
\affiliation{Departamento de F\'{\i}sica, Instituto de F\'{\i}sica Aplicada (INFAP), Universidad Nacional de San Luis - CONICET, Ej\'ercito de Los Andes 950, D5700HHW, San Luis, Argentina}

\begin{abstract}

Monte Carlo simulations and finite-size scaling analysis have been performed to study the jamming and percolation behavior of linear $k$-mers (also known as rods or needles) on the two-dimensional triangular lattice, considering an isotropic RSA process on a lattice of linear dimension $L$ and periodic boundary conditions.
Extensive numerical work has been done to extend previous studies to larger system sizes and longer $k$-mers, which enables the confirmation of a nonmonotonic size dependence of the percolation threshold and the estimation of a maximum value of $k$ from which percolation would no longer occurs. 
Finally, a complete analysis of critical exponents and universality have been done, showing that the percolation phase transition involved in the system is not affected, having the same universality class of the ordinary random percolation.

\end{abstract}


\maketitle

\section{Introduction} \label{introduccion}

Adsorption of extended objects is currently a very active field of research in physics, chemistry and biology. Deposition processes in which the relaxation over typical observation times is negligible can be studied as random sequential adsorption (RSA). In RSA processes particles are randomly, sequentially and irreversibly deposited onto a substrate without overlapping each other. The quantity of interest is the fraction of lattice sites covered at time $t$ by the deposited particles $\theta(t)$. Due to the blocking of the lattice by the already randomly deposited objects, the final state generated by RSA is a disordered state (known as jamming state $\theta_j$), in which no more elements can be deposited due to the absence of free space of appropriate size and shape, $\theta_j \equiv \theta(t \rightarrow \infty)<1$. This phenomenon plays an important role in numerous systems where the deposition process is irreversible over time scales of physical interest \cite{Erban, Evans, Privman, Senger, Talbot,Cadilhe}.

When a fraction $\theta$ of the lattice is covered by particles, nearest-neighbor occupied sites form structures called clusters. If the concentration of the deposited objects is large enough, a cluster extends from one side to the other of the lattice. The minimum concentration of elements for which this phenomenon occurs is named the percolation threshold $\theta_p$, and determines a phase transition in the system \cite{Zallen,Stauffer,Sahimi,Grimmett,Bollo}. As discussed in previous paragraph, $\theta$ ranges from 0 to $\theta_j$ for objects occupying more than one site, and the interplay between jamming and percolation must be considered.

Despite the simplicity of its definition, it is well-known that it is a quite difficult matter to analytically determine the value of the jamming coverage and percolation threshold. For some special types of lattices, geometrical considerations enable to derive their jamming and percolation thresholds exactly, i.e, one-dimensional (1D) substrates \cite{Redner} and monomers (particles occupying one lattice site) for two-dimensional (2D) systems \cite{Stauffer}.


In the case of lattice models of extended objects deposited on 2D lattices, which is the topic of this paper, the inherent complexity of the system still represents a major difficulty to the development of accurate analytical solutions, and  computer simulations appear as a very important tool for studying this subject. In this direction, several authors investigated the deposition of linear $k$-mers on a two-dimensional (2D) square lattice \cite{Bonnier,Vandewalle,EPJB1,Kondrat}. The results obtained revealed that: (1) the jamming coverage decreases monotonically approaching the asymptotic value of $\theta_j=0.66(1)$ for large values of $k$; (2) the percolation threshold is a nonmonotonic function of the size $k$: it decreases for small rod sizes, goes through a minimum around $k=13$, and finally increases for large segments; and (3) the ratio of the two thresholds $\theta_p/\theta_j$ has a complex behavior: after initial growth, it stabilizes between $k=3$ and $k=7$, and then it grows again.


The RSA problem becomes more difficult to solve when the objects are deposited on a 2D triangular lattice, and only very moderate progress has been reported so far \cite{Budi1,Budi2,Budi3,Budi4}. In the line of present work, Budinski-Petkovi\'c and Kozmidis-Luburi\'c \cite{Budi1} examined the kinetics of the RSA of objects of various shapes on a planar triangular lattice. The coverage of the surface and the jamming limits were calculated by Monte Carlo simulation. In all cases, the authors found that the jamming coverage decreases monotonically as the $k$-mer size increases: $\theta_j= \theta_0+\theta_1 \exp{(-k/r)}$, where $\theta_0$, $\theta_1$ and $r$ are parameters that depend on the shape of the adsorbing object. In the case of straight rigid $k$-mers, the simulations were performed for values of $k$ between 1 and 11 and lattice size $L = 128$.

Later, Budinski-Petkovi\'c et al. \cite{Budi2} investigated percolation and jamming thresholds for RSA of extended objects on triangular lattices. Numerical simulations were performed for lattices with linear size up to $L=1000$, and objects of different sizes and shapes (linear segments; angled objects; triangles and hexagons). It was found that for elongated shapes the percolation threshold monotonically decreases, while for more compact shapes it monotonically increases with the object size. In the case of compact objects such as triangles and hexagons, a no-percolation regime was observed. In the case of linear segments with values of $k$ up to 20, the obtained results revealed that (1) the jamming coverage monotonically decreases with $k$, and tends to 0.56(1) as the length of the rods increases; (2) the percolation threshold decreases for shorter $k$-mers, reaches a value $\theta_p \approx  0.40$ for $k = 12$, and, it seems that $\theta_p$ does not significantly depend on $k$ for larger $k$-mers; and (3) consequently, the ratio $\theta_p/\theta_j$ increases with $k$.

The effects of anisotropy \cite{Budi3} and the presence of defects on the lattice \cite{Budi4} were also studied by the group of Budinski-Petkovi\'c et al. In summary, despite over two decades of intensive work, the current conjectures for the behavior of the percolation threshold and jamming concentration as a function of $k$ are based on simulations for relatively short $k$-mers (up to $k=20$). In this context, the main objective of the present paper is to extend the work of Budinski-Petkovi\'c et al. \cite{Budi1,Budi2,Budi3,Budi4} to larger lattice sizes and longer $k$-mers. For this purpose, extensive numerical simulations (with $2\leq k \leq 256$ and $40 \leq L/k \leq 160$) supplemented by analysis using finite-size scaling theory have been carried out. Our study allows (1) to obtain more accurate values of percolation and jamming thresholds; (2) to improve the predictions on the behavior of the system for long rods; and (3) to perform a complete analysis of critical exponents and universality.

The paper is organized as follows: the model is described in section \ref{modelo}. The kinetics and jamming coverage are studied in section \ref{cinetica}. The percolation properties are presented in section \ref{percolacion}: simulation scheme, section \ref{simulacion}; dependence of the percolation threshold on the size $k$, section \ref{umbral}; and analysis of the critical exponents and universality class, section \ref{universa}. Finally, conclusions are given in section \ref{conclusiones}.

\section{Model} \label{modelo}


Let us consider the substrate represented by a 2D triangular lattice of $M = L \times L$ sites. In the filling process, straight rigid $k$-mers (with $k \geq  2$) are deposited randomly, sequentially and irreversibly on an initially empty lattice. This procedure, known as random sequential adsorption (RSA), is as follows: (i) one of the three $(x_1,x_2,x_3)$ possible lattice directions and a starting site are randomly chosen; (ii) if, beginning at the chosen site, there are $k$ consecutive empty sites along the direction selected in (i), then a $k$-mer is deposited on those sites. Otherwise, the attempt is rejected. When $N$ rods are deposited, the concentration is $\theta= kN/M$. 

In this paper, and in order to efficiently occupy the sites of the lattice, we randomly select empty $k$-uples from the set of empty $k$-uples, instead of from the whole lattice. This strategy improves significantly the computational cost of the algorithm.

\section{Kinetics and jamming coverage}\label{cinetica}

In order to calculate the jamming thresholds, the probability $W_L(\theta)$ that a lattice of linear size $L$ reaches a coverage $\theta$ will be used \cite{PHYSA2015}. In the simulations, the procedure to determine $W_L(\theta)$ consists of the following steps: (a) the construction of an $L-$lattice (initially empty) and (b) the deposition of particles on the lattice up to the jamming limit $\theta_j$. The jamming limit is reached when it is not possible to adsorb any more $k$-mers on the surface. In the late step, the quantity $m_i(\theta)$ is calculated as
\begin{equation}
m_i(\theta)=\left\{
\begin{array}{cc}
1 & {\rm for}\ \ \theta \leq \theta_j \\
0  & {\rm for}\ \ \theta > \theta_j .
\end{array}
\right.
\end{equation}
$n$ runs of such two steps (a)-(b) are carried out for obtaining the number $m(\theta)$ of them for which a lattice  reaches a coverage $\theta$,
\begin{equation}\label{m}
m(\theta) = \sum_{i=1}^n m_i(\theta).
\end{equation}
Then, $W_L(\theta)=m(\theta)/n$ is defined and the procedure is repeated for different values of $L$. A set of $n= 10^5$ independent samples is numerically prepared for several values of the lattice size ($L/k =$ 100, 150, 200, 300). The $L/k$ ratio is kept constant to prevent spurious effects due to the $k$-mer size in comparison with the lattice linear size $L$.

For infinite systems ($L \rightarrow \infty$), $W_L(\theta)$ is a step function, being 1 for $\theta \leq \theta_j$ and 0 for $\theta > \theta_j$. For finite values of $L$, $W_L(\theta)$ varies continuously between 1 and 0, with a sharp fall around $\theta_j$. As shown in Ref. \cite{PHYSA2015}, the jamming coverage can be estimated from the curves of the probabilities $W_L$ plotted versus $\theta$ for several lattice sizes. In the vicinity of the limit coverage, the probabilities show a strong dependence on the system size. However, at the jamming point, the probabilities adopt a nontrivial value $W^*_L$, irrespective of system sizes in the scaling limit. Thus, plotting $W_L(\theta)$ for different linear dimensions $L$ yields an intersection point $W^*_L$, which gives an accurate estimation of the jamming coverage in the infinite system.

\begin{figure}[t]
\centering
\includegraphics[width=0.95\columnwidth,clip=true]{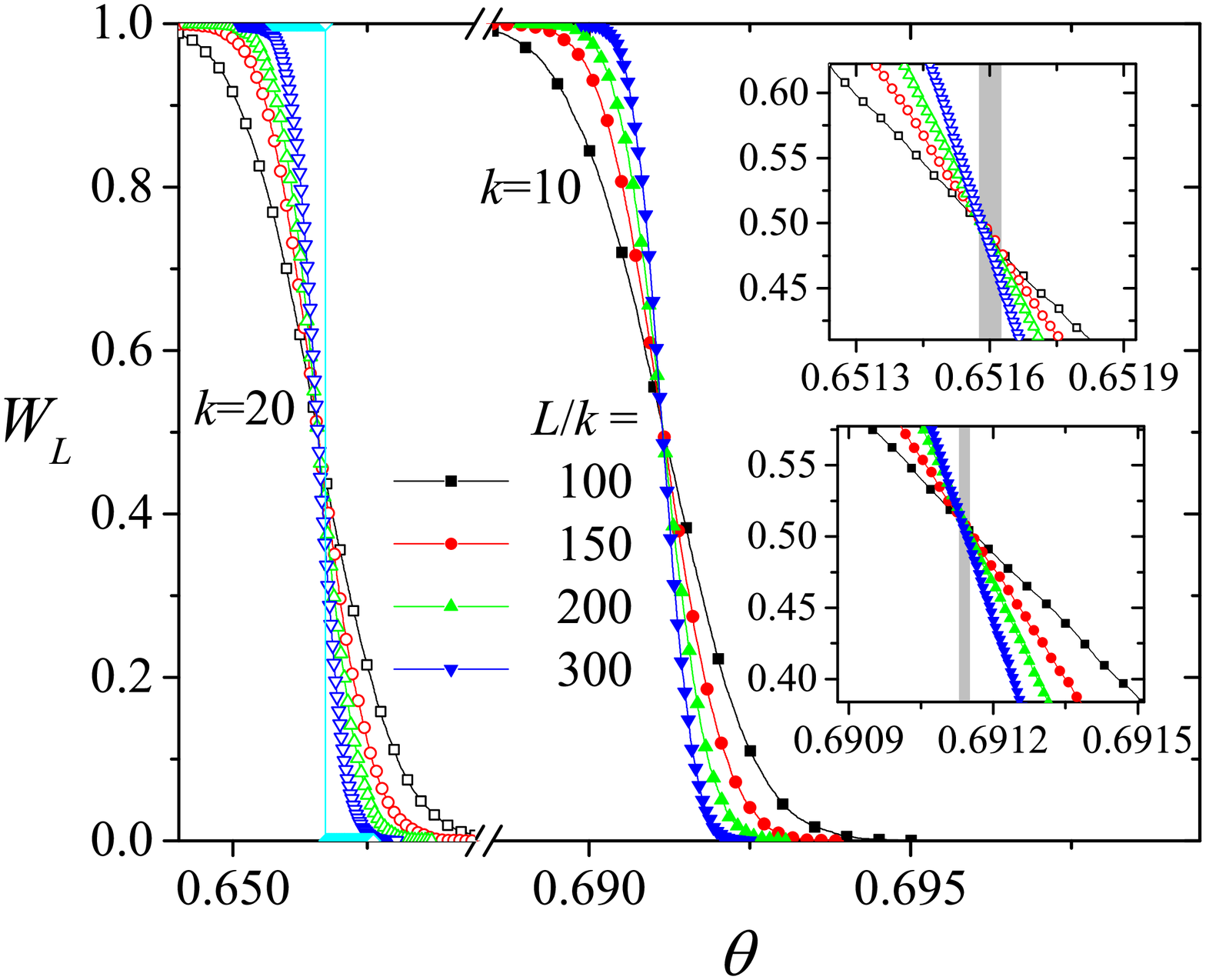}
\caption{\label{fig2} Curves of $W_L$ as a function of the density $\theta$ for several values of $L/k$ (as indicated) and two typical cases, $k=10$ and $k=20$, as indicated. Insets: zoom of the main figure in the vicinity of the intersection points. The grey strip indicates the region where the intersections occur and their width is an estimation of the error.}
\end{figure}

In Fig. \ref{fig2}, the probabilities $W_L(\theta)$ are shown for different values of $L/k$ (as indicated) and two typical cases: (a) $k=10$ (left); and (b) $k=20$ (right). The curves of $W_L(\theta)$ were obtained on a set of $n = 10 ^5$ runs. From the inspection of the figure (and from data do not shown here for a sake of clarity), it can be seen that: (a) for each $k$, the curves cross each other in a unique point $W^*_L$; (b) those points do not modify their numerical value for the different cases studied, being $W^*_L \approx 0.50$; (c) those points are located at very well defined values in the $\theta$-axes determining the jamming threshold for each $k$, $\theta_{j,k}$; and (d) $\theta_{j,k}$ decreases for increasing values of $k$.

\begin{figure}[t]
\centering
\includegraphics[width=0.95\columnwidth,clip=true]{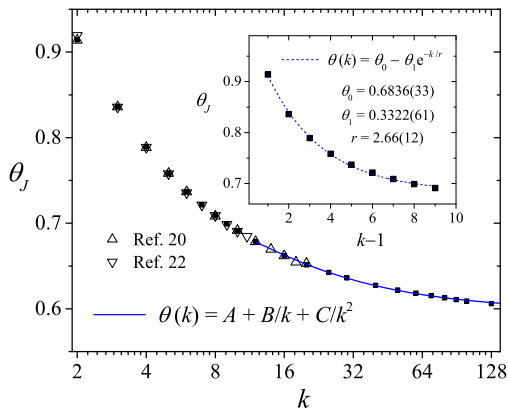}
\caption{ \label{fig3}
Jamming coverage $\theta_{j,k}$ as a function of $k$ for linear $k$-mers on triangular lattices with $k$ between 2 and 128. Inset: As main figure for $2 \leq k \leq 10$. Solid squares represent simulation results (second column of Table I), open symbols denote previous data in the literature \cite{Budi2,Budi4}, and lines correspond to the fitting functions as discussed in the text.}
\end{figure}

The procedure of ~{Fig. \ref{fig2}} was repeated for $k$ ranging between 2 and 128. The results are shown in Fig. \ref{fig3} and compiled in the second column of Table I. Two well-differentiated regimes can be observed. In the range $2 \leq k \leq 20$, the values obtained of $\theta_{j}$ coincide with those reported in Refs. \cite{Budi2} and \cite{Budi4}, and can be fitted with the function proposed in Ref. \cite{Budi1}: $\theta_{j,k}= \theta_0+\theta_1 \exp{(-k/r)}$, with $\theta_0=0.684(3)$, $\theta_1=0.332(6)$ and $r=2.66(2)$ (see inset). These results validate our program and calculation method.

For large values of $k$, the data follow a similar behavior to that predicted by Bonnier et al. \cite{Bonnier} for square lattices: $\theta_{j,k}= A + B/k + C/k^2$ ($k \geq 12$), being $A=\theta_{j,k=\infty}= 0.5976(5)$ the result for the limit coverage of a triangular lattice by infinitely long $k$-mers, $B=1.268(30)$ and $C=-3.61(34)$. 

The value $\theta_{j,k=\infty}= 0.5976(5)$ improves the previously obtained in Ref. \cite{Budi2} using an exponential fit, showing the advantages of having reached larger sizes for the objects.

\section{Percolation}\label{percolacion}

\subsection{Simulation scheme}\label{simulacion}

As it was already mentioned, the central idea of percolation theory is based on finding the minimum concentration $\theta=\theta_p$ for which a cluster extends from one side of the system to the opposite. We are interested in determining \textit{i)} the dependence of $\theta_p$ as a function of the size $k$, and \textit{ii)} the universality class of the phase transition occurring in the system.

The finite-scaling theory gives us the basis to determine the percolation threshold and the critical exponents of a system with a reasonable accuracy. For this purpose, the probability $R=R^{X}_{L,k}(\theta)$ that an $L-$lattice percolates at the concentration $\theta$ of occupied sites by rods of size \textit{k} can be defined
\cite{Stauffer,Binder,Yone1}. Here, the following definitions can be given according to the meaning of $X$:
\begin{itemize}
  \item $R^{x_1}_{L,k}(\theta)$: the probability of finding a percolating cluster along the $x_1$-direction,\\
  \item $R^{x_2}_{L,k}(\theta)$: the probability of finding a percolating cluster along the $x_2$-direction,\\
  \item $R^{x_3}_{L,k}(\theta)$: the probability of finding a percolating cluster along the $x_3$-direction,.
  \end{itemize}
Other useful definitions for the finite-size analysis are:
\begin{itemize}
  \item $R^{U}_{L,k}(\theta)$: the probability of finding a cluster which percolates on any direction,\\
  \item $R^{I}_{L,k}(\theta)$: the probability of finding a cluster which percolates in the three $(x_1,x_2,x_3)$ directions,\\
  \item $R^{A}_{L,k}(\theta)$=$\frac{1}{3}[R^{x_1}_{L,k}(\theta)+R^{x_2}_{L,k}(\theta)+R^{x_3}_{L,k}(\theta)]$.
\end{itemize}

Computational simulations were applied to determine each of the previously mentioned quantities. Each simulation run consists of the following steps: (a) the construction of a triangular lattice of linear size $L$ and coverage $\theta$, (b) the cluster analysis using the Hoshen and Kopelman algorithm \cite{Hoshen}. In the last step, the size of largest cluster $S_L$ is determined, as well as the existence of a percolating island.

A total of $m_{L}$ independent runs of such two steps procedure were carried out for each lattice size $L$. From these runs a number $m^X_{L}$ of them present a percolating cluster, this is done for the desired criterion among $X = {x_1,x_2,x_3, I, U,A}$. Then, $R^X_{L,k}(\theta)= m^X_{L} / m_{L}$ is defined and the procedure is repeated for different values of $L$, $\theta$ and $k$.

In addition to the different probabilities  $R^X_{L,k}(\theta)$, the percolation order parameter $P$ and the corresponding susceptibility $\chi$ have been measured \cite{Biswas,Chandra},
\begin{equation}\label{parord}
P=\langle S_{L}\rangle/M,
\end{equation}
and
\begin{equation}\label{chi}
\chi=[\langle S_{L} ^2\rangle-\langle
S_{L}\rangle ^2]/M,
\end{equation}
where $S_{L}$ represents the size of the largest cluster and $\langle ... \rangle$ means an average over simulation runs.

In our percolation simulations, we used $m_{L}= 10^5$. In addition, for each value of $\theta$, the effect of finite size was investigated by examining square lattices with $L/k =$ 32, 40, 50, 75, 100. As it can be appreciated this represents extensive calculations from the numeric point of view. From there on, the finite-scaling theory can be used to determine the percolation threshold and the critical exponents with a reasonable accuracy.

\begingroup
\squeezetable
\begin{table}[h]
\label{T1}
\begin{center}
\caption{Jamming coverage versus $k$. The values marked with * have been digitized from Fig. 4 of Ref. \cite{Budi2}.}
\begin{tabular}{|p{1cm} p{2cm} p{2cm} p{2cm}|}
\hline 
$k$ &$\theta_J$ &$\theta_J$ (Ref. \cite{Budi2})  &$\theta_J$ (Ref. \cite{Budi4}) \\ 
\hline 

2	& 0.9142(12)	& 0.9139(5)	& 0.9194(5)\\
3	& 0.8364(6)		& 0.8362(7)	& 0.8358(5)\\
4	& 0.7892(5)		& 0.7886(8)	& 0.7888(7)\\
5	& 0.7584(6)		& 0.758 *	& 0.7579(6)\\
6	& 0.7371(7)		& 0.737 *	& 0.7356(8)\\
8	& 0.7091(6)		& 0.708 *	& 0.7089(8)\\
10	& 0.6912(6)		& 0.692 *	& 0.6906(9)\\
12	& 0.6786(6)		& 0.678 *	&			\\
20	& 0.6515(6)		& 0.653 *	&			\\
30	& 0.6362(6)		& 			& 			\\
40	& 0.6276(6)		& 			& 			\\
50	& 0.6220(7)		& 			& 			\\
60	& 0.6183(6)		& 			& 			\\
70	& 0.6153(6)		& 			& 			\\
80	& 0.6129(7)		& 			& 			\\
90	& 0.6108(7)		& 			& 			\\
100	& 0.6090(8)		& 			& 			\\
128	& 0.6060(13)	& 			& 			\\

\hline 
\end{tabular} 
\end{center}
\end{table}
\endgroup

\subsection{Percolation threshold}\label{umbral}

\begin{figure}[t]
\centering
\includegraphics[width=0.95\columnwidth,clip=true]{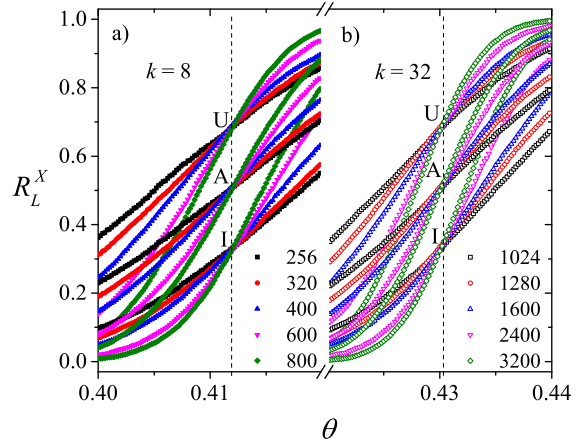}
\caption{ \label{fig4} Fraction of percolating lattices $R^X_{L,k}(\theta)$ ($X= I, U, A$ as indicated) as a function of the concentration $\theta$ for $k = 8$ (a), $k = 32$ (b) and different lattice sizes: $L/k = 32$, squares; $L/k = 40$, circles; $L/k = 50$, up triangles; $L/k = 75$, down triangles; and $L/k = 100$, diamonds. Vertical dashed line denotes the percolation threshold $\theta_{p,k}$ in the thermodynamic limit.}
\end{figure}

The standard theory of finite-size scaling \cite{Stauffer,Binder,Yone1} allows for various efficient routes to estimate the percolation threshold from simulation data. One of these methods, which will used here, is from the curves of $R^X_{L,k}(\theta)$.

In Fig. \ref{fig4}, the probabilities $R^I_{L,k}(\theta)$, $R^U_{L,k}(\theta)$ and $R^A_{L,k}(\theta)$ are presented for two typical cases: (a) $k=8$ (left); and (b) $k=32$ (right). In order to express these curves as a function of continuous values of $\theta$, it is convenient to fit $R^{X}_{L,k}(\theta)$ with some approximating function through the least-squares method. The fitting curve is the {\em error function} because $dR^{X}_{L,k}(\theta)/d\theta$ is expected to behave like the Gaussian distribution ~\cite{Yone1}
\begin{equation}\label{ecu1}
    \frac{dR^{X}_{L,k}}{d\theta}=\frac{1}{\sqrt{2\pi}\Delta^{X}_{L,k}}\exp \left\{ -\frac{1}{2} \left[\frac{\theta-\theta_{p,k}^{X}(L)}{\Delta^{X}_{L,k}}
    \right] \right\},
\end{equation}
where $\theta_{p,k}^{X}(L)$ is the concentration at which the slope of $R^{X}_{L,k}(\theta)$ is the largest and $\Delta^{X}_{L,k}$ is the standard deviation from $\theta_{p,k}^{X}(L)$.

Once obtained the values of $\theta_{p,k}^{X}(L)$ for different lattice sizes, a scaling analysis can be done \cite{Stauffer}. Thus, we have
\begin{equation}
\theta_{p,k}^{X}(L)= \theta_{p,k}^{X}(\infty) + A^X L^{-1/\nu},
\label{extrapolation}
\end{equation}
where $A^X$ is a non-universal constant and $\nu$ is the critical exponent of the correlation length which will be taken as 4/3 for the present analysis, since, as it will be shown in Subsec. \ref{universa}, our model belongs to the same universality class as random percolation \cite{Stauffer}.

\begin{figure}[t]
\centering
\includegraphics[width=0.95\columnwidth,clip=false]{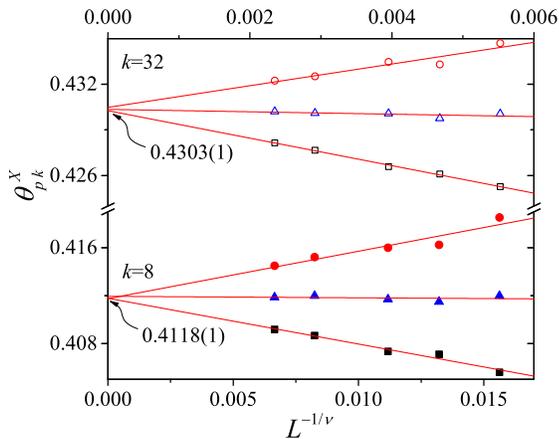}
\caption{\label{fig5}Extrapolation of $\theta_{p,k}^{X}(L)$ towards the thermodynamic limit according to the theoretical prediction given by Eq. (\ref{extrapolation}). Circles, triangles and squares denote the values of $\theta_{p,k}^{X}(L)$ obtained by using the criteria I, A and U, respectively. Different values of $k$ are presented: (a) $k=8$ and (b) $k=32$.}
\end{figure}

Fig. \ref{fig5} shows the plots towards the thermodynamic limit of $\theta_{p,k}^{X}(L)$ according to Eq. (\ref{extrapolation}) for the data in Fig. 4. From extrapolations it is possible to obtain $\theta_{p,k}^{X}(\infty)$ for the criteria $I$, $A$ and $U$. Combining the three estimates for each case, the final values of $\theta_{p,k}(\infty)$ can be obtained. Additionally, the maximum of the differences between $|\theta_{p,k}^{U}(\infty)-\theta_{p,k}^{A}(\infty)|$ and $|\theta_{p,k}^{I}(\infty)-\theta_{p,k}^{A}(\infty)|$ gives the error bar for each determination of $\theta_{p,k}(\infty)$. In this case, the values obtained were: $\theta_{p,k=8}(\infty)=0.4118(1)$ and $\theta_{p,k=32}(\infty)=0.4303(1)$. For the rest of the paper, we will denote the percolation threshold for each size $k$ by $\theta_{p,k}$ [for simplicity we will drop the ``$(\infty)$"].

\begingroup
\squeezetable
\begin{table}[b]
\label{T2}
\begin{center}
\caption{Percolation threshold versus $k$. The values marked with * have been digitized from Fig. \ref{fig4} of Ref. \cite{Budi2}.}
\begin{tabular}{|p{1cm} p{2cm} p{2cm}|}
\hline 
$k$ &$\theta_p$ &$\theta_p$ (Ref. \cite{Budi2}) \\ 
\hline 

2	& 0.4876(5)	& 0.4841(13)	\\
4	& 0.4449(13)	& 0.4399(12)	\\
8	& 0.4118(1)	& 0.407 *	\\
12	& 0.4092(5)	& 0.400 *	\\
16	& 0.4124(6)	& 0.406 *	\\
20	& 0.4169(3)	& 0.401 *	\\
32	& 0.4303(1)	& 	\\
64	& 0.4523(4)	& 	\\
80	& 0.4597(3)	& 	\\
128	& 0.4737(8)	& 	\\
192	& 0.4844(5)	& 	\\
256	& 0.4887(7)	& 	\\

\hline 
\end{tabular} 
\end{center}
\end{table}
\endgroup

\begin{figure}[t]
\centering
\includegraphics[width=0.95\columnwidth,clip=true]{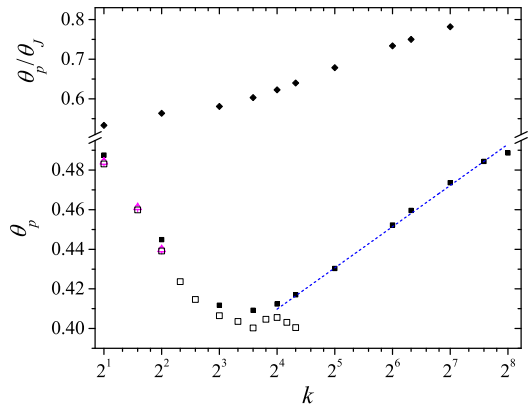}
\caption{\label{fig6}Squares represent the percolation threshold $\theta_{p,k}$ as a function of $k$ for linear $k$-mers on triangular lattices with $k$ between 2 and 256 (second column of Table II). Open symbols denote previous data in the literature \cite{Budi2}. Diamonds represent the ratio $\theta_p/\theta_j$ and dash line corresponds to the the fitting function $\theta_{p,k}= a + b \log{k}$.}
\end{figure}

The procedure of Fig. \ref{fig5} was repeated for $k$ ranging between 2 and 256, and the results are shown in Fig. \ref{fig6} (solid squares) and collected in the second column of Table II. A nonmonotonic size dependence is observed for the percolation threshold, which decreases for small particles sizes, goes through a minimum around $k = 13$, and finally grows for large segments. 
This striking behavior has already been observed for the percolation threshold of $k$-mers on square lattice \cite{Bonnier, Kondrat, Tara1}, and can be interpreted as a consequence of the local alignement effects occurring for larger $k$ (long needles) and their influence on the structure of the critical clusters \cite{Bonnier, Tara1}.

We tried to fit the obtained data for larger $k$ ($k = 16 ... 256$), using the function $\theta_{p,k}= a + b \log{k}$, being $a=0.3265(26)$ and $b=0.03003(70)$.

In Fig. \ref{fig6} can also be observed the ratio of percolation and jamming concentrations, $\theta_p/\theta_j$, which shows a monotonically increasing behavior. Combining the fitting functions used for both concentrations we obtain an estimation for this ratio which increases, for large $k$, proportionally to $\log{k}$. In this way, the condition $\theta_p/\theta_j \simeq 1$ corresponds to a value of $k \simeq 10^4$ from which percolation would no longer occur, in accordance with previous observations in the square geometry for rods \cite{Tara1} and specially in the case of $k \times k$ squares \cite{Nakamura}.

\begin{figure}[t]
\centering
\includegraphics[width=0.49\columnwidth,trim=25 0 110 60,clip=true]{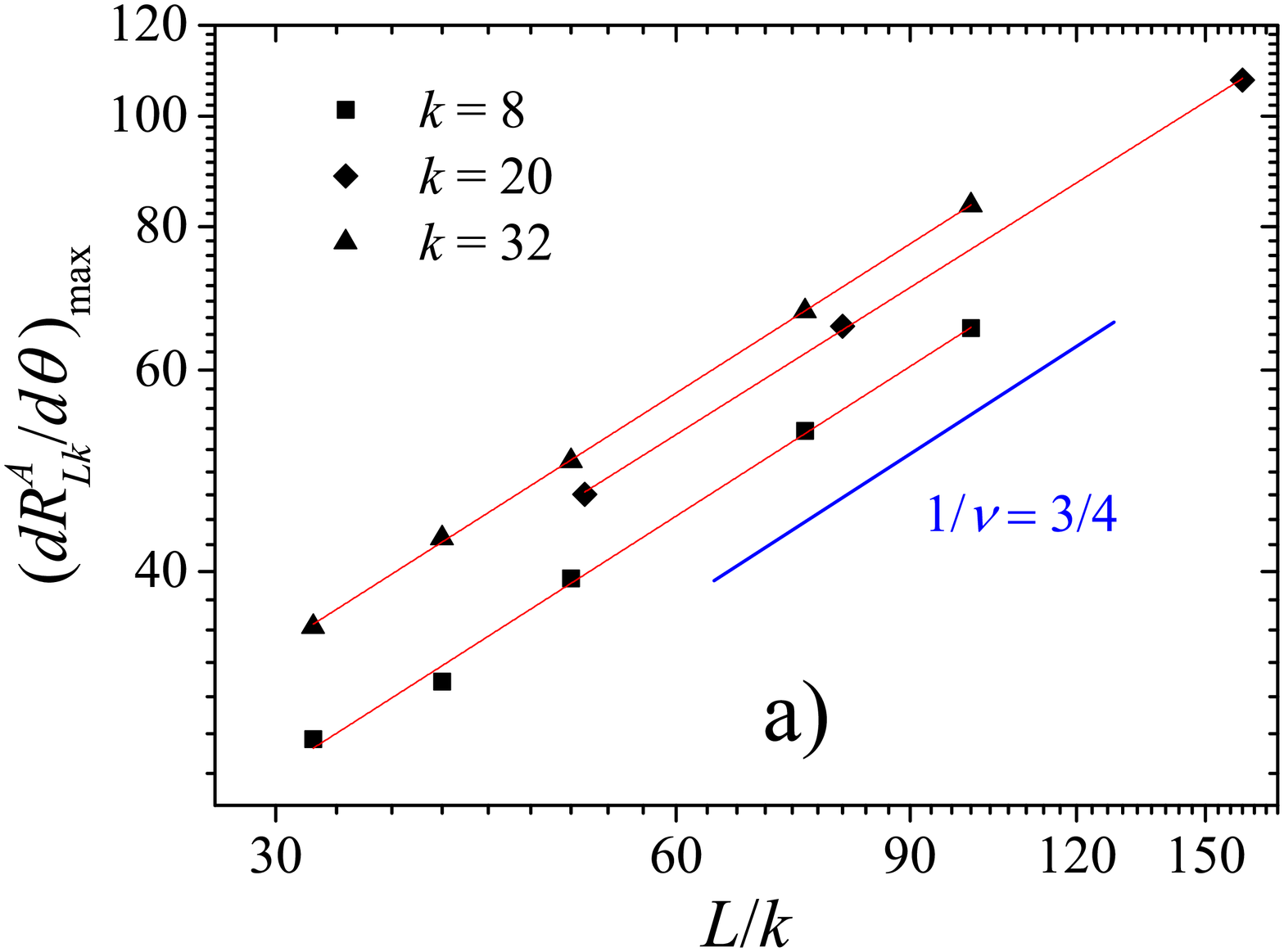}
\includegraphics[width=0.49\columnwidth,trim=25 0 110 60,clip=true]{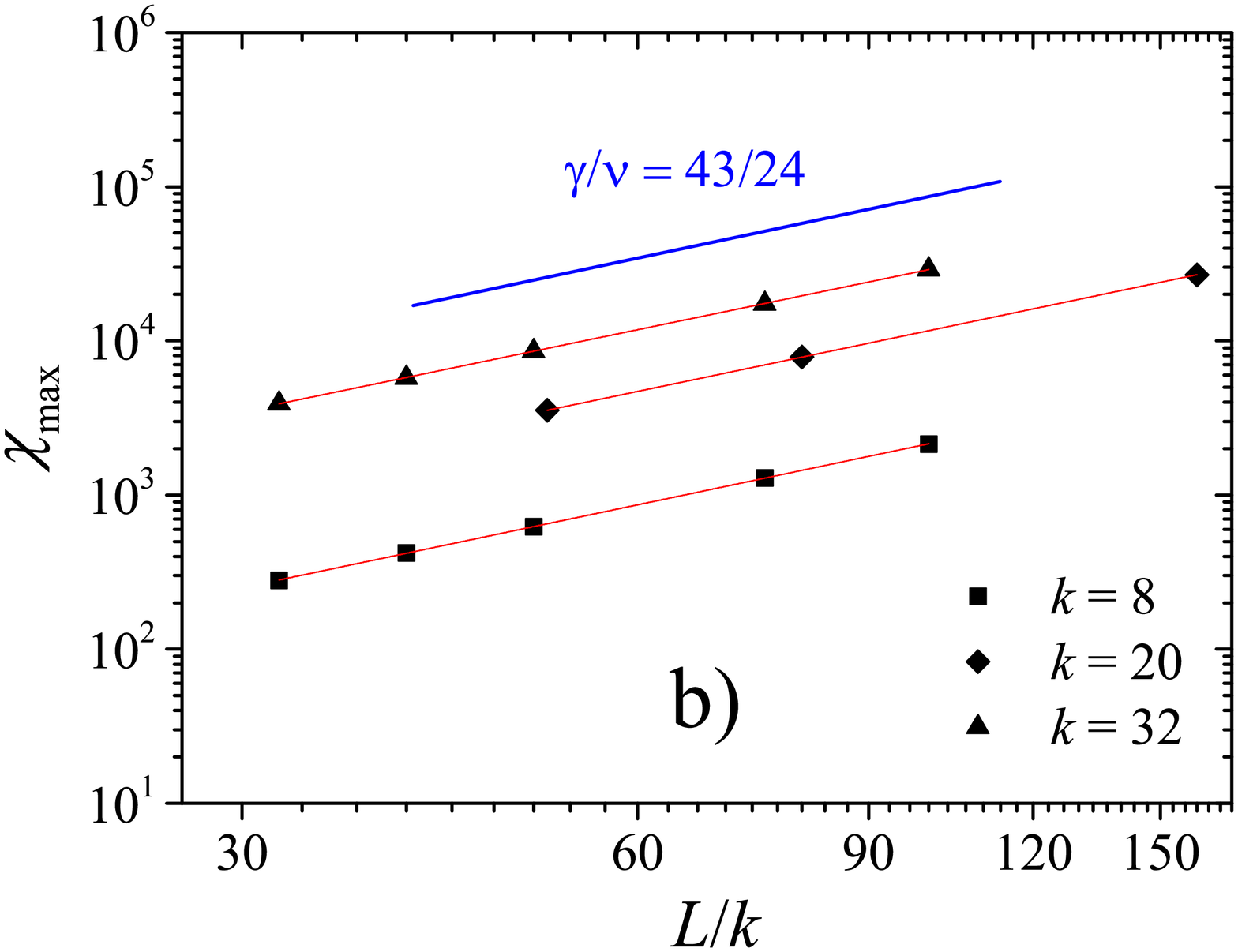}
\includegraphics[width=0.49\columnwidth,trim=25 0 110 60,clip=true]{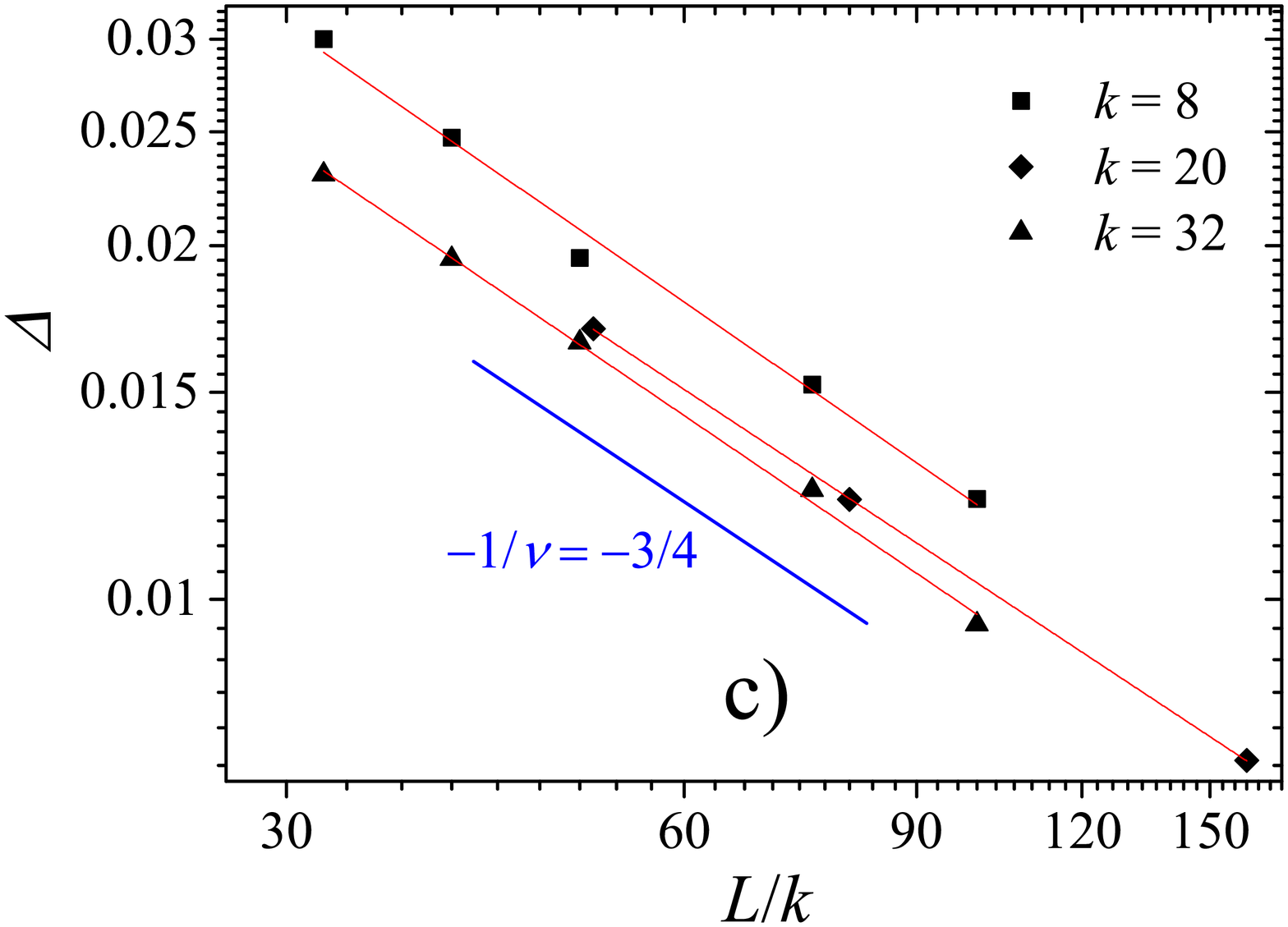}
\includegraphics[width=0.49\columnwidth,trim=25 0 110 60,clip=true]{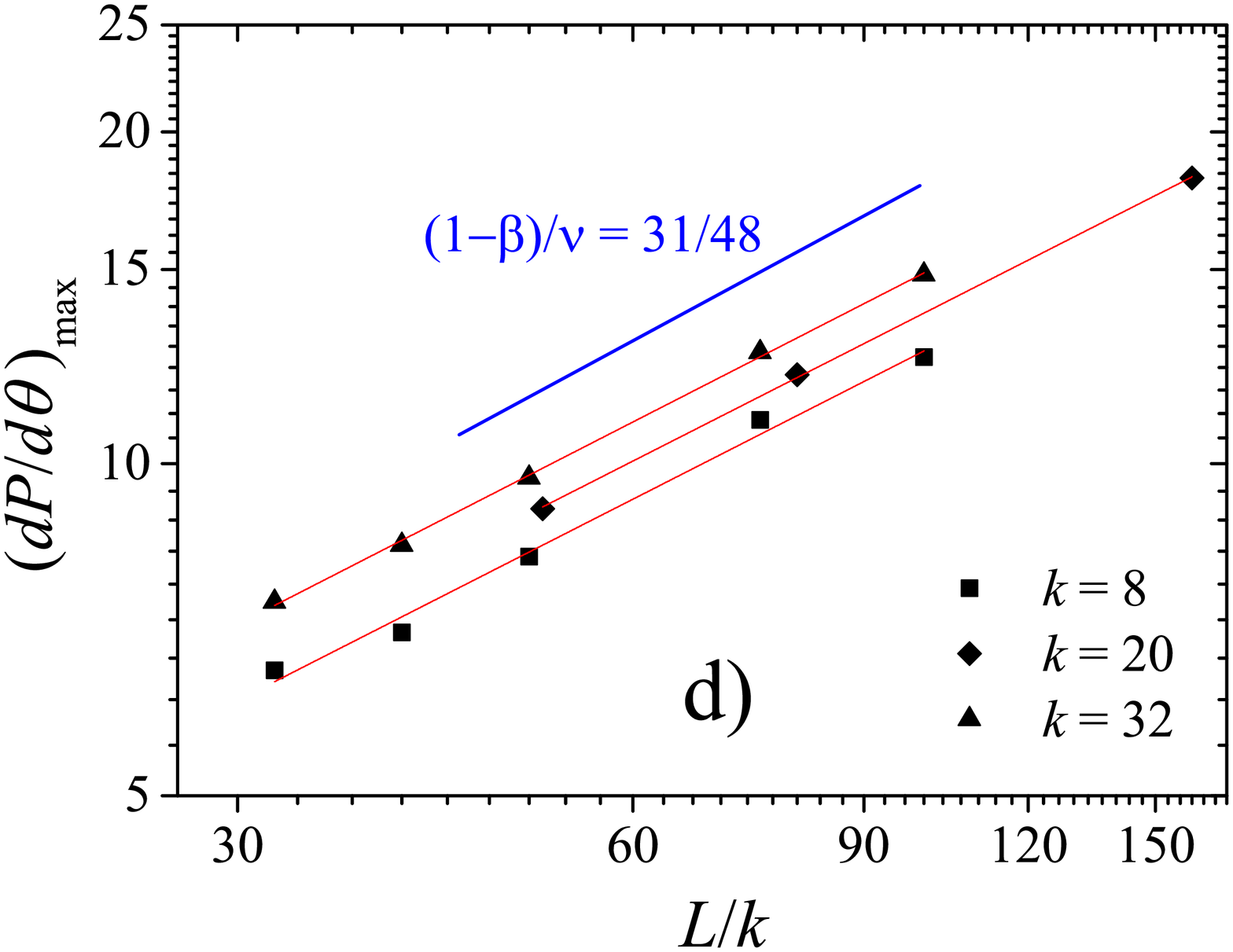}
\caption{\label{fig7}(a) Log-log plot of $\left(d R^{A}_{L,k}/d \theta \right)_{\rm max}$ as a function of $L/k$ for $k=X$ (solid circles) and $k=X$ (open circles). According to Eq. (\ref{lambda}) the slope of each line corresponds to $1/ \nu=3/4$. (b) Log-log plot of $\chi_{\rm max}$ as a function of $L/k$ for $k=X$ (solid circles) and $k=X$ (open circles). The slope of each line corresponds to $\gamma/ \nu=43/24$. (c) Log-log plot of $\left(dP/d\theta\right)_{\rm max}$ as a function of $L/k$ for $k=X$ (solid circles) and $k=5$ (open circles). According to Eq. (\ref{functionPmax}), the slope of each curve corresponds to $(1-\beta)/\nu=31/48$.}
\end{figure}

\subsection{Critical exponents and universality class}\label{universa}

In this section, the critical exponents $\nu$, $\beta$ and $\gamma$  will be calculated. Critical exponents are of importance because they describe the universality class of a system and allow for the understanding of the related phenomena.

The standard theory of finite-size scaling allows for various methods to estimate $\nu$ from numerical data. One of these methods is from the maximum of the function in Eq. (\ref{ecu1}) \cite{Stauffer},
\begin{equation}
\left(\frac{d R^{X}_{L,k}}{d \theta} \right)_{\rm max} \propto L^{1/\nu}. \label{lambda}
\end{equation}

In Fig. \ref{fig7}(a), $\ln\left[\left(d R^{A}_{L,k}/d \theta \right)_{\rm max}\right]$ has been plotted as a function of $\ln\left[ L \right]$ (note the log-log functional dependence) for $k=8$, $k=20$ and $k=32$. According to Eq. (\ref{lambda}) the slope of each line corresponds to $1/ \nu$. As it can be observed, the slopes of the curves remain constant (and close to 3/4) for all studied cases. Thus, $\nu=1.36(3)$ for $k=8$; and  $\nu=1.35(2)$ for $k=32$. The results coincide, within numerical errors, with the exact value of the critical exponent of the ordinary percolation $\nu=4/3$.

Once we know $\nu$, the exponent $\gamma$ can be determined by scaling the maximum value of the susceptibility Eq. (\ref{chi}). According to the finite-size scaling theory \cite{Stauffer}, the behavior of $\chi$ at criticality is $\chi=L^{\gamma/\nu} \overline{\chi}\left( u \right)$, where $u=\left( \theta - \theta_{p,k} \right) L^{1/\nu}$ and $\overline{\chi}$ is the corresponding scaling function. At the point where $\chi$ is maximal, $u=$const. and $\chi_{\rm max} \propto L^{\gamma/\nu}$. Our data for $\chi_{\rm max}$ are shown in Fig. \ref{fig7}(b). The values obtained are  $\gamma=2.35(1)$ for $k=8$ and $\gamma=2.38(1)$ for $k=32$. Simulation data are consistent with the exact value of the critical exponent of the ordinary percolation, $\gamma=43/18$.

On the other hand, the standard way to extract the exponent ratio $\beta$ is to study the scaling behavior of $P$ at criticality \cite{Stauffer},
\begin{equation}
P=L^{-\beta/\nu} \overline{P}\left( u' \right), \label{functionP}
\end{equation}
where $u'=| \theta - \theta_{p,k} | L^{1/\nu}$ and  $\overline{P}$ is the scaling function. At the point where $dP/d\theta$ is maximal, $u=$const. and
\begin{equation}
\left(\frac{dP}{d\theta}\right)_{\rm max}=L^{(-\beta/\nu+1/\nu)} \overline{P}\left( u' \right) \propto L^{(1-\beta)/\nu}. \label{functionPmax}
\end{equation}

The scaling of $(dP/d\theta)_{\rm max}$ is shown in Fig. \ref{fig7}(c). From the slopes of the curves, the following values of $\beta$ were obtained: $\beta=0.18(2)$ for $k=8$ and $\beta=0.19(4)$ for $k=32$. These results agree very well with the exact value of $\beta$ for ordinary percolation, $\beta=5/36=0.14$.

The protocol described in Fig. \ref{fig7} was repeated for $k$ between 2 and 128. In all cases, the values obtained for $\nu$, $\gamma$ and $\beta$ clearly indicate that, independently of the size $k$, this problem belongs to the same universality class that the random percolation.

The scaling behavior can be further tested by plotting $R^{X}_{L,k}(\theta)$ vs $\left(\theta - \theta_{p,k} \right)L^{1/\nu}$, $PL^{\beta/\nu}$ vs $| \theta - \theta_{p,k} |L^{1/\nu}$ and $\chi L^{-\gamma/\nu}$ vs $\left(\theta - \theta_{p,k} \right)L^{1/\nu}$ and looking for data collapsing \cite{Stauffer} (see supplementary material \cite{Supple}).


\section{Conclusions}
\label{conclusiones}

In this paper, extensive numerical simulations and finite-size scaling theory have been used to study the percolation properties of straight rigid rods of length $k$ out of equilibrium (RSA adsorption) as well as the jamming threshold on the two-dimensional triangular lattice.

A nonmonotonic size dependence was found for the percolation threshold $\theta_{p}$, which decreases for small particles sizes in accordance with previous data in the literature \cite{Budi2}. Moreover, for values of $k > 13$ we observe an increasing value of $\theta_{p}$. This striking behavior, observed also in $k$-mers on square lattice, is related to local alignement effects that affects the structures of the critical clusters \cite{Bonnier, Tara1}. The interplay between the percolation and the jamming effects suggests the existence of a maximum value of $k$ from which percolation no longer occurs. 

Finally we observe that the percolation phase transition occurring in the system is not affected, having the same universality class of the ordinary random percolation.

\section*{Acknowledgements}

This work was supported in part by CONICET (Argentina) under project number PIP 112-201101-00615; Universidad Nacional de San Luis (Argentina) under project 322000; and the National Agency of Scientific and Technological Promotion (Argentina) under project  PICT-2013-1678. The numerical work were done using the BACO parallel cluster (composed by  50 PCs each with an Intel i7-3370 / 2600 processor) located  at Instituto de F\'{\i}sica Aplicada, Universidad Nacional de San Luis - CONICET, San Luis, Argentina.

\renewcommand{\baselinestretch}{1.0}

\end{document}


\widetext

\begin{center}
\textbf{\large Supplementary Material for:}
\end{center}

\begin{center}
\textbf{\large Jamming and percolation in random sequential adsorption of straight rigid rods on a two-dimensional triangular lattice}
\end{center}

\setcounter{figure}{0}

\section*{Data collapsing}
The scaling behavior can be further tested by plotting $R^{X}_{L,k}(\theta)$ vs $\left(\theta - \theta_{p,k} \right)L^{1/\nu}$, $PL^{\beta/\nu}$ vs $| \theta - \theta_{p,k} |L^{1/\nu}$ and $\chi L^{-\gamma/\nu}$ vs $\left(\theta - \theta_{p,k} \right)L^{1/\nu}$ and looking for data collapsing \cite{Stauffersup}. Figs. 1 and 2 show the excellent collapse of curves of $R^A_{L,k}$ (a), $P$ (b), $\chi$ (c) and the cumulant $U_L$ (d), for two typical cases ($k=8$ and $k=32$) and different lattice sizes, as indicated. The plots were made using the value of $\theta_{p,k=8}=0.4118$ and $\theta_{p,k=32}=0.4303$ calculated above and the exact values of the critical exponents of the ordinary percolation $\nu=4/3$, $\beta=5/36$ and $\gamma=43/18$. This leads to independent controls and consistency checks of the values of all the critical exponents.

\begin{figure}[h]
\centering
\includegraphics[width=0.45\textwidth, trim=25 0 60 0, clip=true]{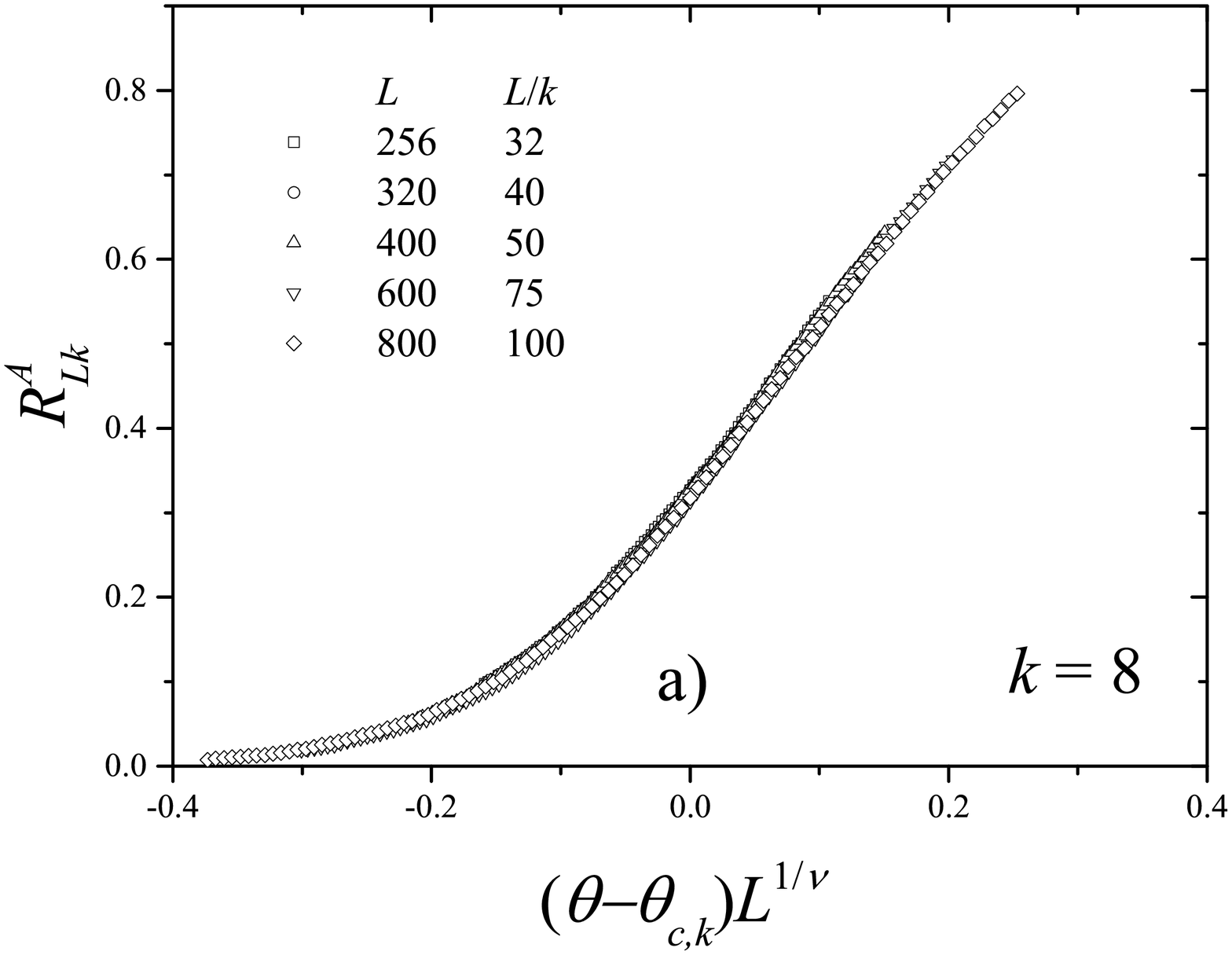}
\includegraphics[width=0.45\textwidth, trim=25 0 60 0, clip=true]{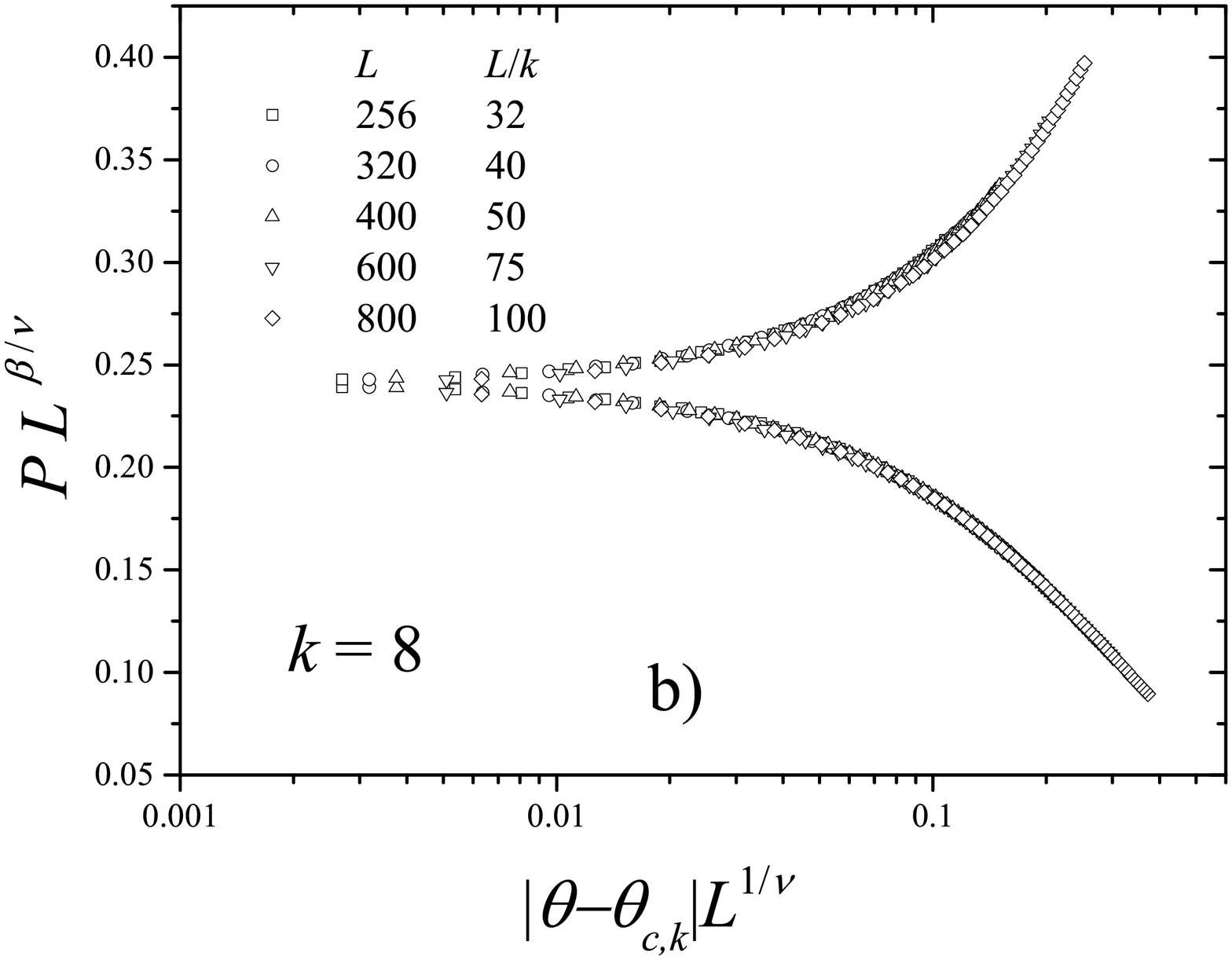}
\includegraphics[width=0.45\textwidth, trim=25 0 60 0, clip=true]{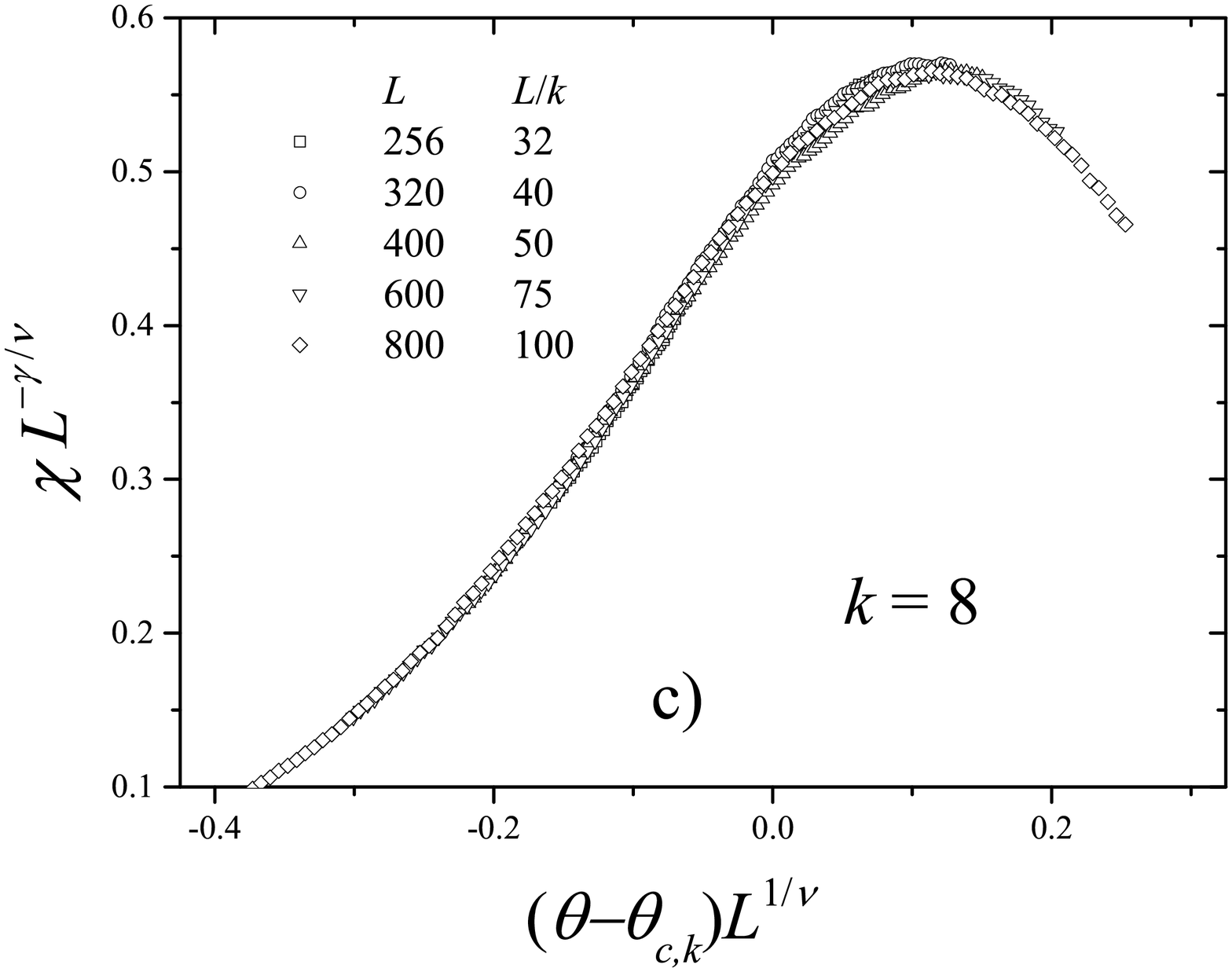}
\includegraphics[width=0.45\textwidth, trim=25 0 60 0, clip=true]{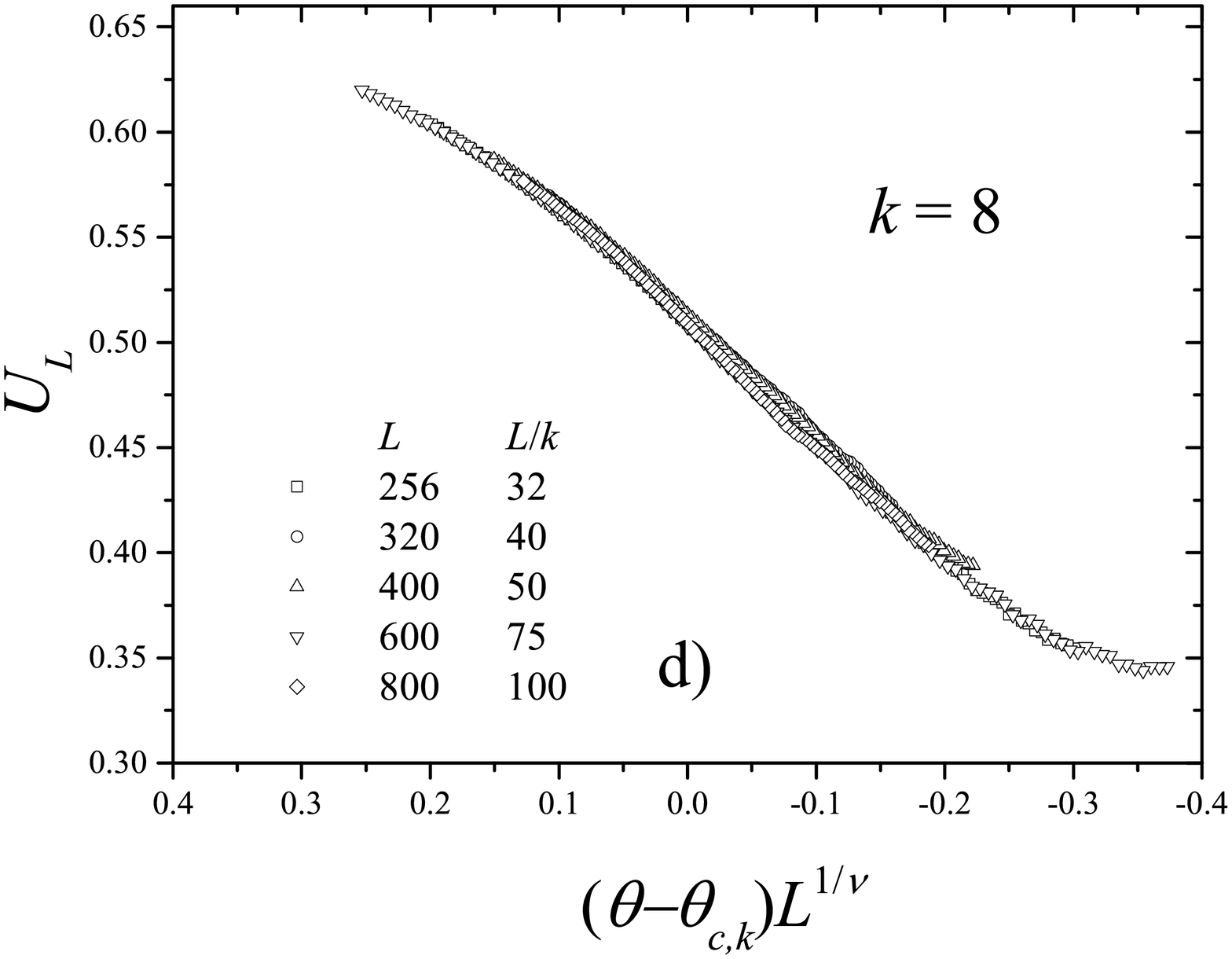}
\label{fig8}
\caption{Data collapsing of the percolation probability, $R^{A}_{L,k}(\theta)$ vs $\left(\theta - \theta_{p,k} \right)L^{1/\nu}$ (a), for $k=8$. Data collapsing of the percolation order parameter, $PL^{\beta / \nu}$ vs $| \theta - \theta_{p,k} | L^{1/\nu}$ (b). Data collapsing of the susceptibility, $\chi L^{-\gamma / \nu}$ vs $(\theta - \theta_{p,k}) L^{1/\nu}$ (c). 
Data collapsing of the percolation cumulant, $U_L$ vs $(\theta - \theta_{p,k}) L^{1/\nu}$ (d). The plots were made using $\theta_{p,k=8}=0.4118$, $\theta_{p,k=32}=0.4303$ and the exact percolation exponents $\nu=4/3$, $\beta=5/ 36$ and $\gamma=43/18$.}
\end{figure}

\begin{figure}[h]
\centering
\includegraphics[width=0.45\textwidth, trim=25 0 60 0, clip=true]{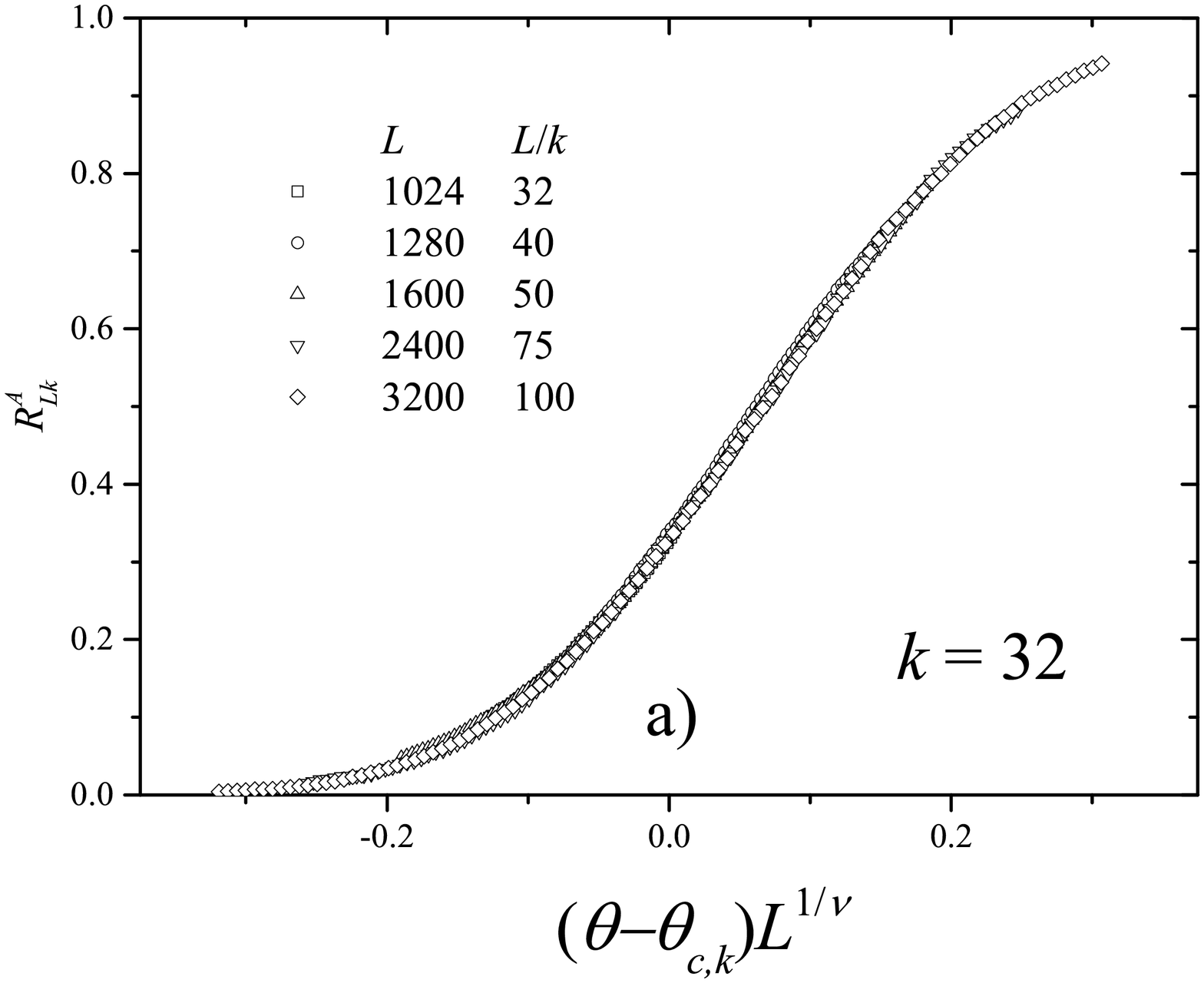}
\includegraphics[width=0.45\textwidth, trim=25 0 60 0, clip=true]{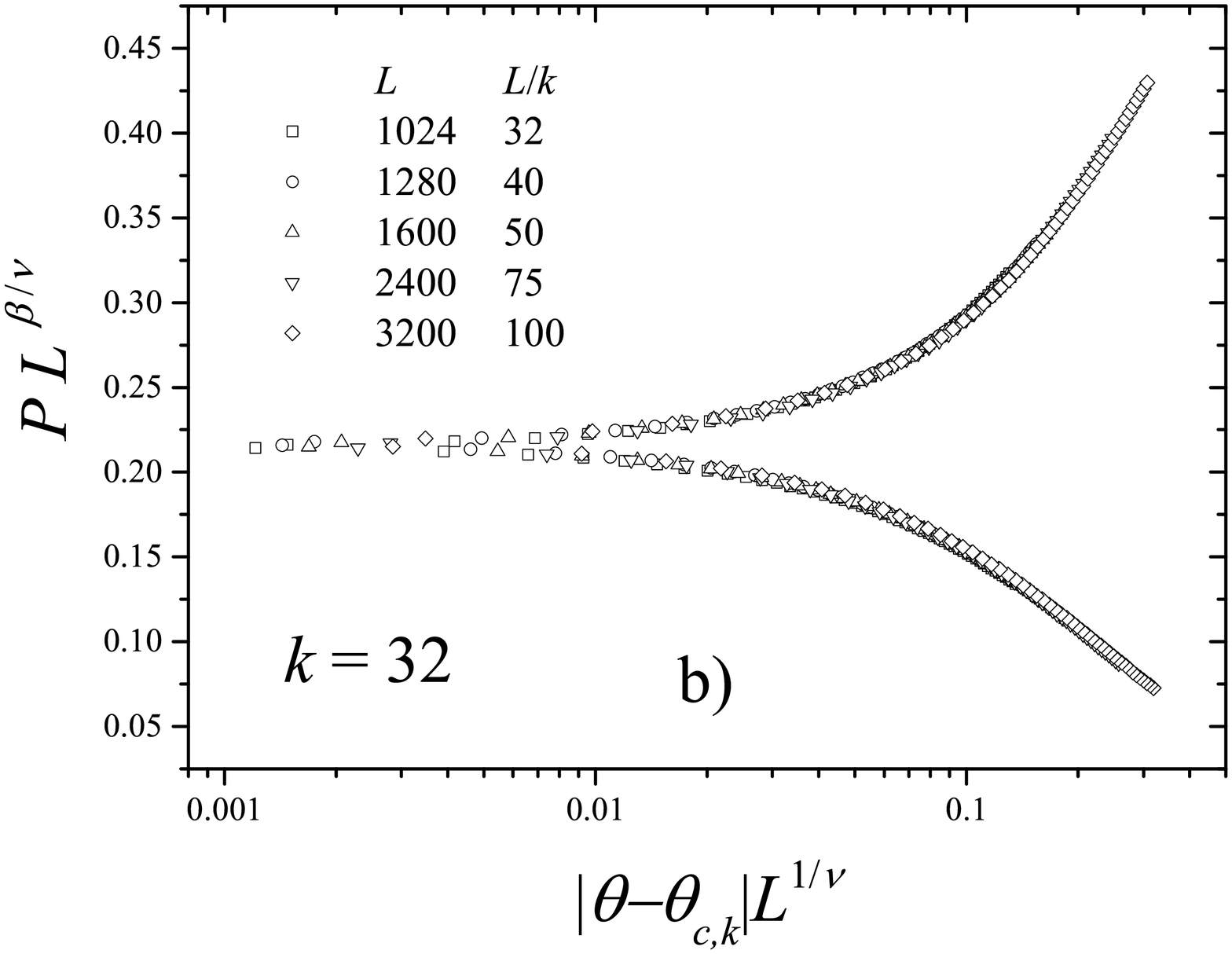}
\includegraphics[width=0.45\textwidth, trim=25 0 60 0, clip=true]{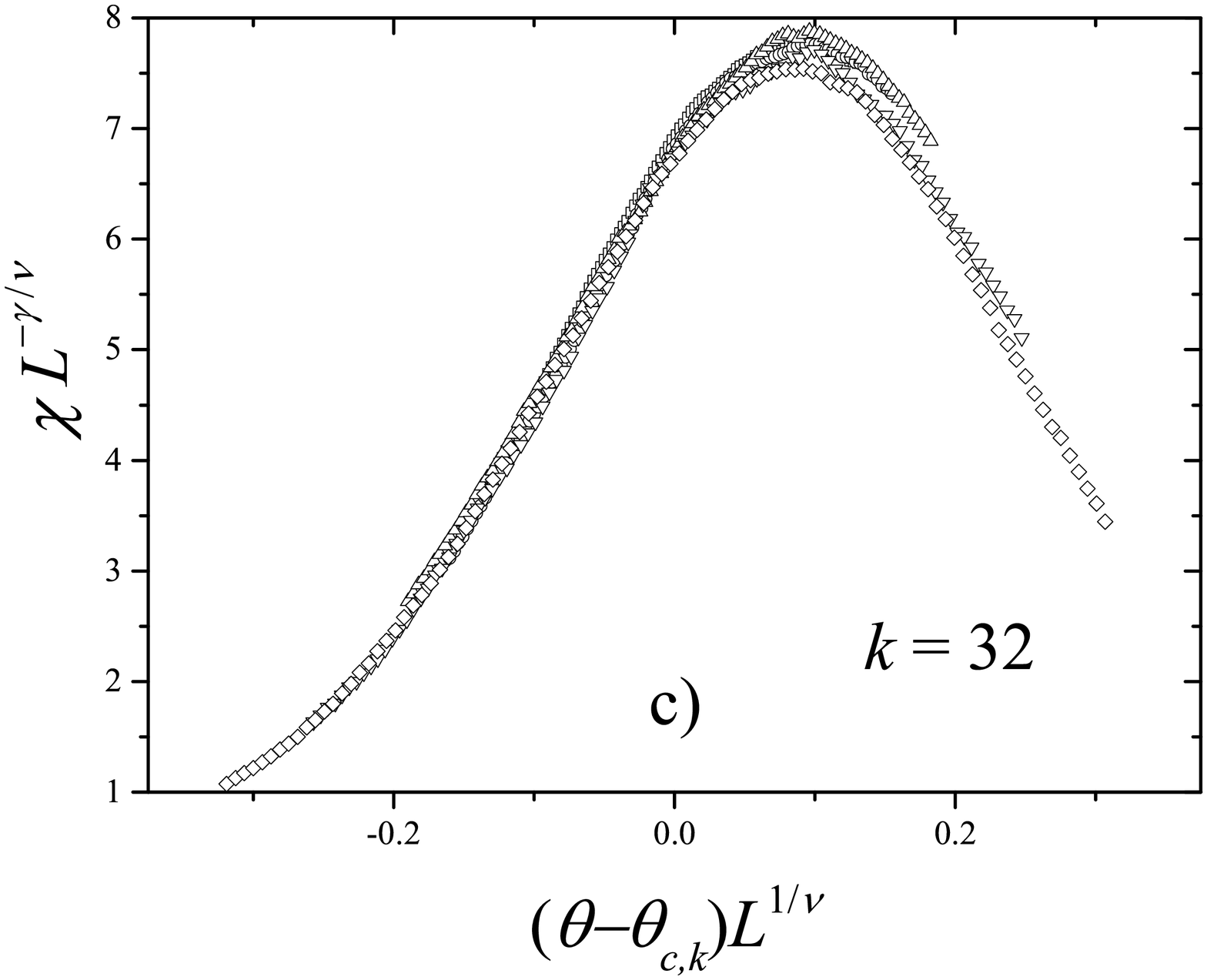}
\includegraphics[width=0.45\textwidth, trim=25 0 60 0, clip=true]{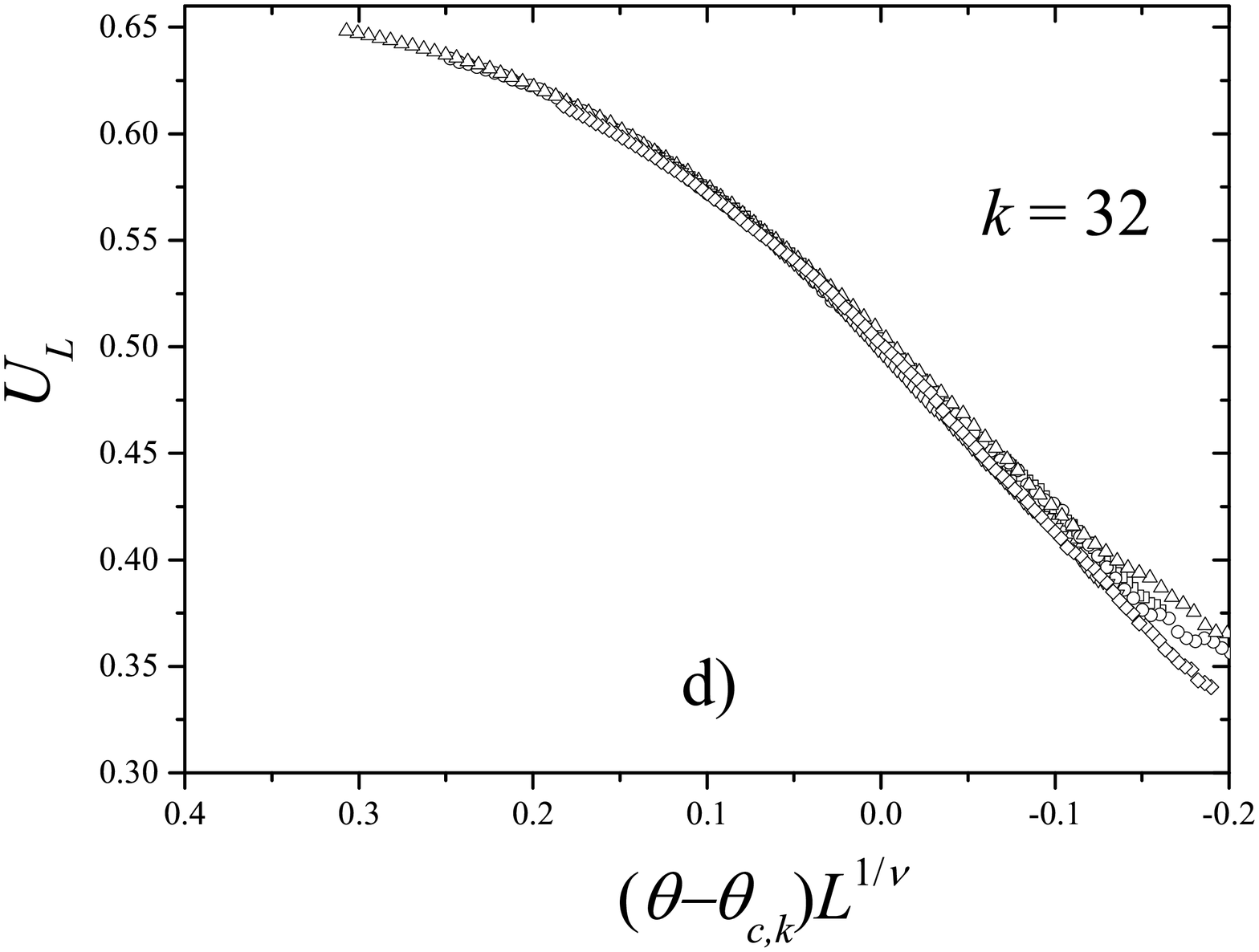}
\label{fig8}
\caption{Idem to Fig. 1 for $k=32$.}
\end{figure}